\documentclass[prb,twocolumn,showpacs,amsmath,amssymb,superscriptaddress]{revtex4}%
\usepackage{graphicx}
\usepackage{dcolumn}
\usepackage{bm}

\newcommand{\Rvec}{{\bf R}}
\newcommand{\qvec}{{\bf q}}
\newcommand{\Qvec}{{\bf Q}}
\newcommand{\kvec}{{\bf k}}
\newcommand{\beq}{\begin{equation}}
\newcommand{\eeq}{\end{equation}}
\newcommand{\beqa}{\begin{eqnarray}}
\newcommand{\eeqa}{\end{eqnarray}}

\begin{document}
\title{Spin canting as a result of the competition between stripes and
spirals in cuprates}
\author{G. Seibold}
\affiliation{Institut F\"ur Physik, BTU Cottbus, PBox 101344, 03013 Cottbus,
Germany}
\author{R.S. Markiewicz}
\affiliation{ Physics Department, Northeastern University, Boston MA
02115, USA}
\affiliation{ISC-CNR and Dipartimento di Fisica, Universit\`a di Roma
``La Sapienza'', P. Aldo Moro 2, 00185 Roma, Italy}
\author{J. Lorenzana}
\affiliation{ISC-CNR and Dipartimento di Fisica, Universit\`a di Roma
``La Sapienza'', P. Aldo Moro 2, 00185 Roma, Italy}

\date{\today}
\begin{abstract}
Based on the extended Hubbard model we calculate the energy
of stripe and spiral ground states. We find that uniform spirals get
favored by a large $t'/t$ ratio but are unstable at small doping
towards stripes and checkerboard textures with spin canting.
The structure of these inhomogeneities also depends on $t'/t$ and the
associated spin currents may induce a small lattice distortion
associated with local dipole moments.   We
discuss a new kind of stripe which appears as a domain wall
of the antiferromagnetic (AF) order parameter with a fractional change 
of the phase of the AF order. For
large $|t'/t|$ spirals can be stabilized under certain conditions 
in the overdoped regime which may explain the elastic incommensurate 
magnetic response recently observed in iron-codoped Bi2201 materials.  
\end{abstract}


\maketitle
\section{Introduction}
The investigation of Hubbard type models with regard to solutions 
exhibiting modulated magnetic order is to a large extent motivated 
by (in)elastic neutron scattering experiments on high-temperature 
superconductors (for an overview cf. Ref. \onlinecite{birg06}).
Especially in the underdoped regime many compounds 
show a pronounced  low energy spin response which is peaked 
away from the antiferromagnetic wave-vector. For overdoped samples
of lanthanum cuprates it has been shown \cite{waki04} that these low energy 
incommensurate spin fluctuations vanish at the same concentration
where superconductivity disappears thus suggesting a close relation
between both phenomena.
  
There are basically three possibilities which can account for the observed
incommensurate spin response. First, the system can be close to a 
magnetic instability which in a random-phase approximation based 
treatment would strongly
enhance the magnetic susceptibiliy at the corresponding wave-vectors.
Moreover, in case of a superconducting (SC) d-wave gap the depletion of
low energy spectral weight induces quasi undamped low energy incommensurate
spin excitations which merge at the antiferromagnetic wave-vector into
the famous resonance peak.~\cite{eres1,eres2,eres3} This scenario is very popular in optimally
doped YBCO, since the experimentally observed energy of the resonance peak 
scales with the SC transition temperature $T_c$ as expected for this
model. 
However, one could also envisage a situation of real symmetry breaking
in the spin channel where (in terms of the above counting) the second 
and third possibilities correspond to ordered phases in the transverse and
longitudinal spin channel, respectively.
Since the charge couples to the square of the magnetic order parameter 
one expects in the latter case also a concomitant charge modulation. 
In lanthanum cuprates (LCO), doped with Ba or codoped  with Nd or Eu, 
~\cite{tra95,fujita02,klauss00} the third scenario
is now well established. In these compounds
the spin response even becomes static
and concomitant charge order appears,
which first was evidenced through the coupling to the lattice \cite{tra95}
but recently  
more directly through soft resonant x-ray scattering.~\cite{abb05, fink09} 
The fact that 
the periodicity of the charge modulation is half that of the spin strongly
suggests that these materials have a so called `striped ground state', 
i.e. quasi one dimensional antiphase domain walls of the antiferromagnetic 
order host the doped charge carriers. 
A polarized neutron scattering study on La$_{1.48}$Nd$_{0.4}$Sr$_{0.12}$CuO$_4$\cite{christensen07}
in fact suggests that the magnetic order is one-dimensionally modulated
although some two-dimensional noncollinear structure \cite{fine07}
may not be excluded.
 
It is worth noting that stripe textures have been predicted as stable 
Hartree-Fock (HF) saddle points of the Hubbard model \cite{mach89,schulz90,
zaanen89} before they were found experimentally in cuprates (and nickelates).
However, it was realized early on that one has to include correlations
beyond HF in order to correctly describe the experimental data 
in lanthanum cuprates. In the past
years two of us have performed detailed investigations within the
Gutzwiller approximation (GA) supplemented with Gaussian fluctuations
which allowed for the explanation of the doping dependent incommensurability
and various transport properties \cite{lor02,sei09}, the optical conductivity
\cite{lor03} and magnetic excitations \cite{sei05,sei06} on the basis
of striped ground states. 

Non codoped lanthanum cuprates show a low energy inelastic incommensurability
with a doping dependence rather similar to that of the codoped compounds 
suggesting the existence of some form of fluctuating and (or) disordered 
stripe order.  Moreover, below a doping $\delta=0.055$ the magnetic 
response rotates from vertical (i.e. along the Cu-O bond direction) to 
diagonal, becoming static and one-dimensional with the associated
modulation along the orthorombic $b^*$-axis, also in agreement with the
stripe picture. On the other hand, direct probes of charge order are more
difficult to obtain, leaving some room for the existence of
incommensurate magnetic structures without charge order such as spirals.
Although there are good hints from local probes like NMR, 
NQR \cite{nmr1,nmr2,nmr3,nmr4} and tunneling\cite{kohsaka}
that point to charge ordered states, it is worth examining the
possibility that transverse spin (spiral) textures exist. 

Even more controversial is the situation in YBCO, the second class
of high-T$_c$ compounds which have been intensively investigated by
neutron scattering (NS) experiments.~\cite{birg06} 
Since in optimally doped YBCO the
low energy incommensurate magnetic response shows up below T$_c$, 
one line of thought is that the signal corresponds to the
dispersion of a bound exciton which is formed inside a d-wave
superconducting gap.~\cite{res1,res2,res3,res4,res5}
However, it should be noted that for underdoped
YBCO the observation of an incommensurate 
(and even static) spin response
also  above T$_c$ \cite{hayden04,hinkov07} raises questions about 
this interpretation. 
In any case, the lack of experimental evidence for
charge modulation in YBCO (which if dynamic or disordered is of course
hard to detect)  
makes it difficult to conclusively attribute the incommensurate magnetic
response to stripes as in codoped LCO, though it has been successfully
described within a model based on slowly fluctuating (or disordered) charge
stripes.~\cite{vojta06} Also the quantum oscillations found
in underdoped YBCO \cite{doiron07} are compatible with a reconstruction
of the Fermi surface due to stripe order.~\cite{millis07,Sebastian}

Due to the lack of large single crystals other cuprate superconductors
are less intensively studied by NS than the aforementioned
compounds. However, a recent study has revealed elastic incommensurate
magnetic peaks in an overdoped Bi$_{1.75}$Pb$_{0.35}$Sr$_{1.9}$CuO$_{6+z}$
sample [(Bi,Pb)2201] codoped with iron.~\cite{hiraka10} Within the error 
bars the doping $\delta \approx 0.23$ and 
the measured incommensurability $\varepsilon\approx 0.21$ 
seem to extend the relation 
$\varepsilon \approx \delta$ (which holds for underdoped lanthanum cuprates
\cite{yamada98}) to large doping without saturation at $\delta\approx 0.12$.

An alternative to stripes in order to account for the incommensurate 
spin response in cuprates is based on the formation of spirals.
These textures are characterized by a  homogeneous periodic and 
planar rotation of the Cu moments with no associated charge modulation.
Thus at first glance spirals seem to be a promising candidate for those 
cuprates in which incommensurate magnetic fluctuations have been observed without charge
fluctuations, and corresponding theories have been recently put forward in Refs.
\onlinecite{sush05,sush07,sush09}. 

Within Hubbard-type models 
spiral solutions have been investigated on the basis of 
HF,~\cite{arri91,assad93,chub95}
slave-boson (or GA) \cite{arri91,fres92,assad93,rac06,rac07} 
and dynamical mean-field 
\cite{fleck99} calculations.
The energies of stripes and spirals have been computed
within HF \cite{zaanen96} and slave-boson methods \cite{rac06,rac07}
based on the Hubbard model. These latter investigations have found 
a strong influence of the ratio between next-nearest and nearest
neighbor hopping $t'/t$ on the respective stability of both textures.
Here our starting point is similar, however, we show that
the low doping phase separation instability of spirals can lead 
to the formation of nanoscale charge and spin inhomogeneities
with substantial spin canting. Stripes and spirals 
should therefore be viewed only as limiting cases of these more complex
spin and charge textures. We discuss our findings
with regard to recent neutron scattering experiments on
iron-codoped lanthanum \cite{he10} and bismuth \cite{hiraka10}
cuprate superconductors.
The paper is organized as follows. In Sec. II we introduce the formalism and
present our results in Sec. III. Discussion and conclusions are presented 
in Sec. IV.
\section{Formalism}
Our starting point is the one-band Hubbard model
\begin{equation}\label{HM}
H= \sum_{i,j,\sigma} t_{ij} c_{i,\sigma}^{\dagger}c_{j,\sigma} + U\sum_{i}
n_{i,\uparrow}n_{i,\downarrow},
\end{equation} 
where $c_{i,\sigma}$ ($c^\dagger_{i,\sigma}$) destroys (creates) an electron
with spin $\sigma$ at site
$i$, and $n_{i,\sigma}=c_{i,\sigma}^{\dagger}c_{i,\sigma}$. $U$ is the
on-site Hubbard repulsion and $t_{ij}$ denotes the hopping parameter between
sites $i$ and $j$. We restrict to hopping between nearest ($\sim t$) and
next-nearest ($\sim t'$)neighbors.
Our approach is based on the Gutzwiller variational wave
function $|\Psi_g \rangle =P_g |SD\rangle$ where  $P_g$ is the Gutzwiller
projector and $|SD\rangle$ a Slater determinant. For $|SD\rangle$ we
use a state with arbitrary charge and spin order, including spin
canting. We define the associated one-body density as 
$\rho_{ij}^{\sigma_1,\sigma_2}=\langle
SD|\hat c_{j\sigma_2}^\dagger \hat c_{i\sigma_1}  |SD\rangle$. 
The wave-function optimization problem leads to 
a generalized Gutzwiller approximation\cite{geb} 
which on the saddle-point level is equivalent to the Kotliar-Ruckenstein
slave-boson approach \cite{kr}. The derivation of the  
spin-rotational invariant Gutzwiller energy functional can be found
 in Ref. \onlinecite{jose06}
\begin{equation}\label{ega}
E^{GA}= \sum_{i,j}
t_{ij} \langle{\bf \Psi_i}^\dagger {\bf z_{i}}
{\bf z_{j}}{\bf \Psi_j}\rangle + U\sum_{i} D_i  .
\end{equation}
Here we have defined the spinor operators
\begin{displaymath}
{\bf \Psi_i}^\dagger = (c_{i\uparrow}^\dagger , c_{i\downarrow}^\dagger)
\,\,\,\,\,\,  {\bf \Psi_i} = \left(\begin{array}{c} c_{i\uparrow} \\
c_{i\downarrow} \end{array}\right)
\end{displaymath}
and  the ${\bf z}$-matrix
\begin{eqnarray*}
&&{\bf z}_i= \left( \begin{array}{cc}
z_{i\uparrow}\cos^2\frac{\varphi_i}{2}+z_{i\downarrow}
\sin^2\frac{\varphi_i}{2} &
\frac{S_i^-}{2S_i^z}[z_{i\uparrow}-z_{i\downarrow}]\cos\varphi_i \\
\frac{S_i^+}{2S_i^z}[z_{i\uparrow}-z_{i\downarrow}]\cos\varphi_i&
z_{i\uparrow}\sin^2\frac{\varphi_i}{2}+z_{i\downarrow}\cos^2\frac{\varphi_i}{2}
\end{array} \right) \\
&&\tan^2\varphi_i=\frac{S_i^+S_i^-}{(S_i^z)^2}.
\end{eqnarray*}
and for clarity spin expectation values in the {\em Slater determinant} 
are denoted by
$S_i^+=\rho_{ii}^{\uparrow,\downarrow}$, 
$S_i^-=\rho_{ii}^{\downarrow,\uparrow}$, 
$S_i^z=(\rho_{ii}^{\uparrow,\uparrow}-\rho_{ii}^{\downarrow,\downarrow})/2$,
and 
$\rho_{ii}=\rho_{ii}^{\uparrow,\uparrow}+\rho_{ii}^{\downarrow,\downarrow}$.
In the limit of a vanishing rotation angle $\varphi$ the  ${\bf z}$-matrix
becomes diagonal and the renormalization factors
\begin{widetext}
\begin{displaymath}
z_{i\sigma} = \frac{\sqrt{(1-\rho_i+D_i)(\frac{1}{2}\rho_i+\frac{S_i^z}
{\cos(\varphi_i)}-D_i)}+\sqrt{D_i(\frac{1}{2}\rho_i-\frac{S_i^z}
{\cos(\varphi_i)}-D_i)}}
{\sqrt{(\frac{1}{2}\rho_i+\frac{S_i^z}{\cos(\varphi_i)})(1-\frac{1}{2}\rho_i-
\frac{S_i^z}{\cos(\varphi_i)})}}
\end{displaymath}
\end{widetext}
reduce to those of the standard GA.
Spiral solutions are then computed by minimizing $E^{GA}$ with respect to
a homogeneous
rotation of spins with wave-vector $\Qvec$
\begin{eqnarray}
S_i^x &=& S_0 \cos(\Qvec\Rvec_i) \nonumber \\
S_i^y &=& S_0 \sin(\Qvec\Rvec_i).  \label{eq:spiral}
\end{eqnarray}
In the present paper we usually measure modulations with respect
to AF order $\Qvec_{AF}=(\pi,\pi)$ (lattice constant $a=1$) 
and set $\Qvec=\Qvec_{AF}-\qvec$.  
Stripe solutions are obtained by restricting the magnetization
to the $z$ direction resulting in a  modulation
of $S_i^z$ with $\Qvec$ and a simultaneous modulation
of $\rho_i$ with $2\Qvec$ similar to our previous work
.~\cite{sei04,bob2} However, we also consider stripes and other charge
ordered states with spin canting
in Sec.~\ref{sec:stripes-with-spin}.
\section{Results}
In order to fix the value for the onsite repulsion $U$ we 
refer to a previous paper
where we have shown that a time-dependent extension of
the GA with $U/t=8$ can accurately reproduce 
the magnon excitations of undoped LCO\cite{sei06} as revealed by
neutron scattering.~\cite{col01}
Since $U/t$ should not vary among the cuprate materials
we restrict to $U/t=8$ but investigate the dependence on
the next-nearest neighbor hopping $t'/t$ which from LDA
computations has been shown to specify the various
high-T$_c$ families.~\cite{pava}   
Note that the results of Refs. \onlinecite{rac06,rac07}
have been obtained with a significantly larger value of
$U/t=12$.
  
\subsection{Spirals {\it vs.} stripe states}
\begin{figure}[htb]
\includegraphics[width=8cm,clip=true]{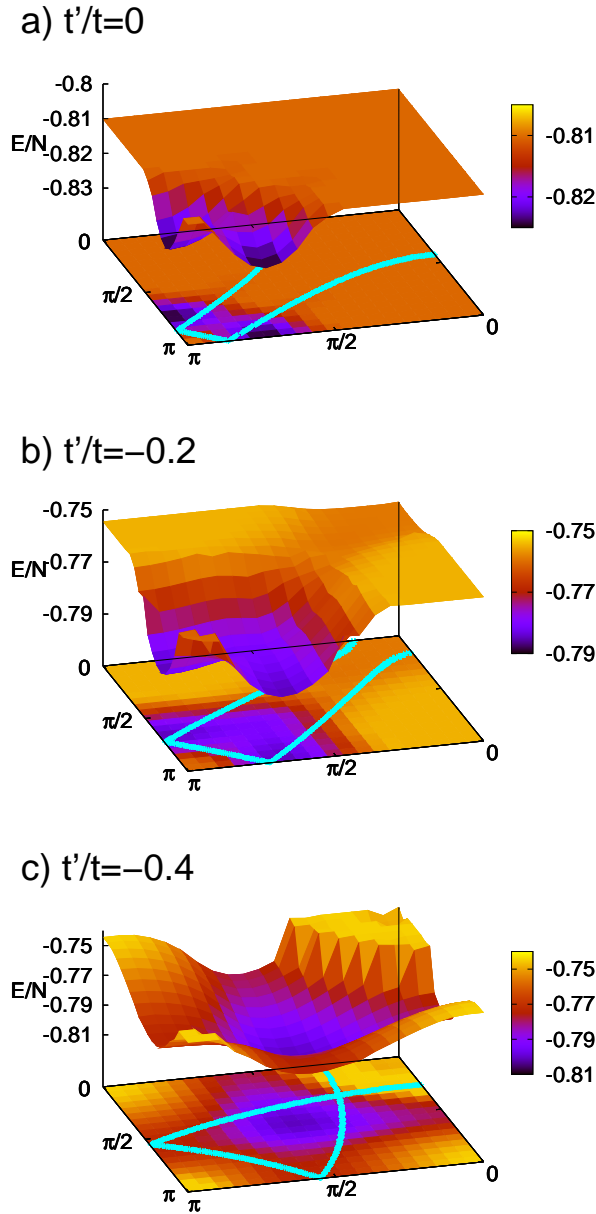}
\caption{(color online) Energy of spiral solutions (per lattice site and
in units of $t$) as a function of spiral momentum
$\Qvec=(q_x,q_y)$ (in units of the lattice constant $a\equiv 1$). 
Doping $\delta=0.2$. The lines in the $(q_x,q_y,0)$-plane correspond to the
nesting curves \cite{bob1}, obtained for the same doping in 
the paramagnetic state.}
\label{fig1}
\end{figure}

Fig. \ref{fig1} shows the energy landscape for spiral solutions
(hole doping measured from half-filling $\delta=0.2$) for different
values of $t'/t$. With increasing $|t'/t|$ one observes a shift
of the minimum from the vertical $(q_x,0)$ and  $(0,q_y)$ directions
towards the diagonal. Thus in the parameter regime relevant for 
LCO compounds $|t'/t|\sim 0.1 \dots 0.2$ the incommensurate scattering 
direction for the corresponding doping 
is correctly reproduced.  

We note that also for other dopings only
vertical or diagonal spiral textures correspond to minima of the
energy landscape in agreement with previous investigations 
.~\cite{arri91,fres92,assad93,rac06,rac07}
This can be easily understood from the nesting curves
\cite{bob1} which are also shown in Fig. \ref{fig1} below the
energy contours. Basically these are ${\qvec}=2 {\kvec}_F$ plots
(${\kvec}_F$ being the Fermi momentum) of the paramagnetic
system backfolded in the $0 \le {\qvec}_{x,y} \le \pi$ quadrant. 
As discussed in Ref. \onlinecite{bob1} the dominant instabilities
occur for those wave vectors which correspond to a crossing of
two nesting curves (`double nesting'). These double nesting points
determine the maximum susceptibilities and are always found to
lie along high-symmetry directions.
The change from vertical to diagonal spirals upon increasing
$t'/t$ can therefore be 
understood from the appearance of an additional `antinodal' double nesting 
vector along the $(0,0) \to (\pi,\pi)$ direction which starts to dominate
over those along the $(\pi,\pi) \to (\pi,0), (0,\pi)$
directions. 

The spiral is a realization of the corresponding
instabilities, but the wave-vectors of energy minima and double nesting
  in Fig. \ref{fig1} do not exactly coincide, as may be seen by comparing
the nesting curves with the contour maps in lower parts of Fig. 
\ref{fig1}a,b,c.  The Stoner criterion gives the 
correct nesting vector
at the instability threshold, but as $U/t$ increases the optimal $q$ can shift. 
From \ref{fig1} it can be seen that for $U/t=8$ the spiral minimum coincides
with the nesting curve near $t'/t=-0.2$, but is closer to (further from) 
$(\pi,\pi)$ for large (smaller) $t'/t$.

\begin{figure}[htb]
\includegraphics[width=8cm,clip=true]{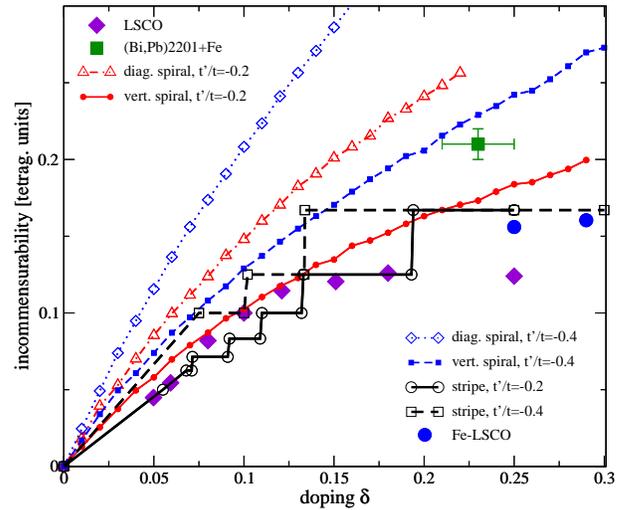}
\caption{(color online) Doping evolution of the spiral and stripe wave-vectors 
$\qvec$ for $t'/t=-0.2$ and 
$t'/t=-0.4$ expressed in terms of the incommensurability. 
Note that the staircase structure
of the stripe incommensurability is only explicitly shown for larger
doping.  The solid diamonds show the doping
dependent incommensurability in LCO (`Yamada plot') and are reproduced
from Ref. \onlinecite{yamada98}.  The solid square corresponds
to the elastic neutron scattering data on iron-codoped (Bi,Pb)2201 
from Ref. \onlinecite{hiraka10}. Solid circles report the incommensurability
of Fe-LSCO from Ref. \onlinecite{he10}. }
\label{fig1b}
\end{figure}

\begin{figure}[htb]
\includegraphics[width=8cm,clip=true]{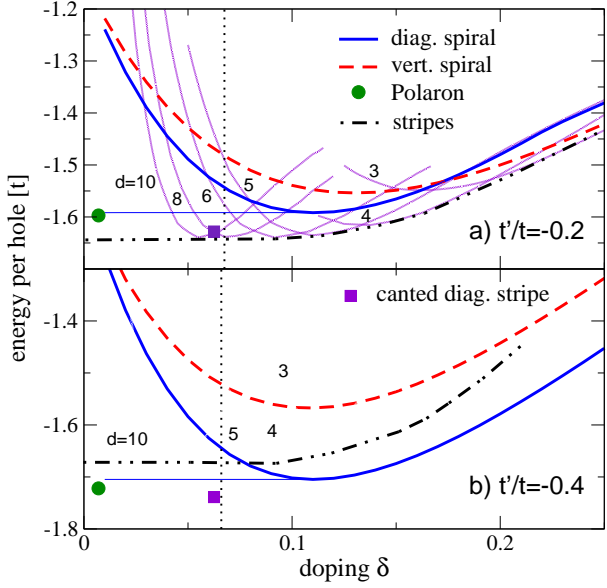}
\caption{(color online) Comparison of diagonal (blue solid) and vertical (red dashed) 
spirals with the energy of stripes (thin grey lines) for $t'/t=-0.2$ (a)
and $t'/t=-0.4$ (b). 
The enveloping curve (dash--dotted) of
the stripe energies is a guide to the eye. Vertical dotted lines indicate the
concentration of zero compressibility for diagonal spirals and the thin solid
line follows from the Maxwell construction. The full circle corresponds
to the energy of a single spin polaron and the square is the energy
of the spin-canted diagonal stripe which is shown in Fig. \ref{fig3}.}
\label{fig2}
\end{figure}

Fig. \ref{fig1b} compares the doping evolution of vertical and
diagonal spiral wave-vectors, for $t'/t=-0.2$ and $t'/t=-0.4$,
with the corresponding values for vertical stripes (cf. also Ref.
\onlinecite{sei04,sei07}).  
The incommensurability is defined in terms of the wave-vector ${\bf
  q}$ of the dominant Fourier component of the magnetization as 
$$\varepsilon=\frac{|\Qvec_{AF}-\qvec|}{2\pi}$$
For stripes we restrict to regular, periodic
stripe ground states with integer $d$ (in terms of the lattice constant)
separation between the charge stripes which amounts to a magnetic
periodicity $2d$ and an incommensurability,
$$\varepsilon=\frac1{2d}$$
which develops a staircase structure, Fig.~2. The steps occur
at the crossing of the  corresponding energy curves
(cf. Fig. \ref{fig2}). It is also convenient to define the number of
doped holes  per unit length along the stripe $\nu=\delta d$ 
which is related to the incommensurability by
\begin{equation}
  \label{eq:epsvsdelta}
\varepsilon=\frac{\delta}{2\nu} .  
\end{equation}

Given a certain number of holes one can ask if it is more favorable to 
add them to a small number of stripes with a large stripe filling or
in a large number of stripes with a small stripe filling.  
For widely separated stripes one can determine an optimum linear density
$\nu_o$ from the minimum of the energy per hole with respect to 
half-filling\cite{lor02,sei04} 
\begin{equation}
e_h=\frac{E_{N_h}-E_{AF}}{N_h} .
\end{equation}
Here $E_{N_h}$ is the total energy of the system doped with $N_h$
holes. We find it convenient to use $e_h$ to characterize the
competition among the different phases at all dopings.  

Fig. \ref{fig2} compares $e_h$ for stripes and spirals as a function of
doping for $t'/t=-0.2$ and $t'/t=-0.4$, respectively.
For stripes the individual  curves correspond to vertically oriented 
domain walls and are labeled by the charge periodicity $d$. The
optimum filling for large $d$ is given by $\nu_o=d\delta_o$
with $\delta_o$ corresponding to the doping at the curve minimum. 
Notice that $\nu_o$ becomes independent of $d$ for large $d$ 
and takes the value 
$\nu_o=0.55 (0.4)$  for $t'/t=-0.2
(-0.4)$. 
The change of $\nu_o$ with $t'$ is expected according to the
results of Ref.~\onlinecite{sei04}.

A constant $\nu_o$ implies via Eq.~\eqref{eq:epsvsdelta} a linear
relation between doping and incommensurability. In other words the
optimum filling determines the slope in the famous Yamada plot.\cite{yamada98} 

Where does the linear relation breaks down? 
The charged core of the stripe has a characteristic width $\xi$ so that
when the charge periodicity $d$ is larger than $\xi$ there are
negligible interstripe interactions and doping proceeds by increasing
the number of stripes.  Therefore the relation is linear for
$d>\xi$. In this regime the value of 
 $e_h$ at the minimum is independent of doping. 

For $d\lesssim
 \xi$ stripe overlap becomes important and doping proceeds by
 increasing the charge of stripes. From Fig.~\ref{fig1b} we see that
 $\xi\sim 4(5)$ for $t'/t=-0.2 (-0.4)$ which can be checked directly
 from the charge profile (see Ref.~\onlinecite{sei04}).

As mentioned above the incommensurability for stripes shows a
staircase structure. For small doping ($d>\xi$), since interstripe interactions
are negligible, one can produce a practically continuous curve by
considering combinations of solutions with 
periodicity $d$ and $d+1$. This becomes very costly for $d\lesssim \xi$ which
is consistent with the tendency of the incommensurability to develop a
plateau at doping larger than $\nu_0 \xi \sim 1/8$.

As remarked before,\cite{lor02,sei04,sei06} 
the linear relation for small doping and the plateau for large doping 
reproduce the doping
dependence of the incommensurate low energy magnetic response of
lanthanum cuprates as revealed by neutron scattering experiments
\cite{yamada98} and reproduced in Fig.~\ref{fig1b}.

For spirals, a minimum of $e_h$ at doping $\delta_s$ indicates
that if the frustration due to the long-range Coulomb interaction is 
ignored \cite{ort06}, the system will lower its energy by phase
separating in undoped regions and spiral regions of doping $\delta_s$.
Contrary to the stripes, for spirals one 
observes a continuous increase of $\varepsilon(\delta)$.  Nevertheless, 
both phases show similar evolution of incommensurability with doping and 
indeed fairly similar values of $\varepsilon(\delta)$ up to $\delta
\sim \nu_o \xi$. 

For larger doping this agreement is lost due to the 
tendency of stripes to make wide plateaus. 
In fact, both spirals and stripes originate from
the same magnetic instability, the doping dependence of which has
been analyzed for cuprate parameters in Ref. \onlinecite{bob1}. In particular,
it is the plateau in the spin susceptibility close to $Q_{AF}=(\pi,\pi)$
(cf. Fig. 1 in Ref. \onlinecite{bob1}) which drives the system unstable 
towards spiral or stripe order for sufficiently large on-site repulsion
$U/t$. We will come back to this point in the next section when discussing
recent elastic neutron scattering experiments on overdoped 
Fe-LSCO \cite{he10} where the resulting incommensurabilities are
shown as solid circles in Fig. \ref{fig1b}.

We proceed by comparing the doping dependence of the energy minima
which determine the orientation and respective stability of 
spirals and stripes.  It should be noted that at low doping diagonal
stripes are almost accidentally degenerate with  vertical stripes with
a negligibly preference for diagonal textures.\cite{sei09} 
 Therefore other mechanisms like long-range
Coulomb interactions, multiband effects, and lattice distortions are
decisive in determining 
the exact stripe state in the very underdoped regime.
In the following we restrict for simplicity to vertical stripes.
The envelope (dashed-dotted
line) for this set of curves represents the doping evolution for the
stripe minimum energy and determines the doping 
dependence of the incommensurability as shown in Fig. \ref{fig1b}.
 
The energy minima of the spiral landscape (cf. Fig. \ref{fig1}) are also 
shown in Fig. \ref{fig2} as  solid  and dashed  lines for diagonal and
vertical spirals, respectively.
As anticipated from Fig. \ref{fig1} the crossover from diagonal
to vertical spirals shifts towards higher doping with increasing $|t'/t|$
and it is found that for $t'/t=-0.4$ diagonal solutions are lower in energy 
over practically the whole doping range.
However, for both diagonal and vertical spirals $e_h$ has a minimum at 
$\delta_s$ which means that for 
doping smaller than $\delta_s$ spiral ground states are unstable with respect
to phase separation as mentioned above.  
The resulting energy of the phase separated solutions
calculated from a Maxwell construction is shown as a thin horizontal 
solid line in Fig. \ref{fig2}.

Also shown in Fig. \ref{fig2} (full circle) is the energy of a single 
hole which is self-trapped in a spin-polaronic state. Since the corresponding
charge is localized within a five-site plaquette with small ferromagnetic
polarization, $e_h$ for a many-polaron state will be independent of
doping up to the point where polaron-polaron interactions become
noticeable.    For $t'/t=-0.2$, both spirals and polarons are unstable
 with respect to the stripe phase, whereas (diagonal) spirals 
become the stable phase for larger $|t'/t|$, consistent with the results of Ref. \onlinecite{bob1}.
  
\subsection{Charge order states  with spin canting}
\label{sec:stripes-with-spin}
\begin{figure}[thb]
\includegraphics[width=9cm,clip=true]{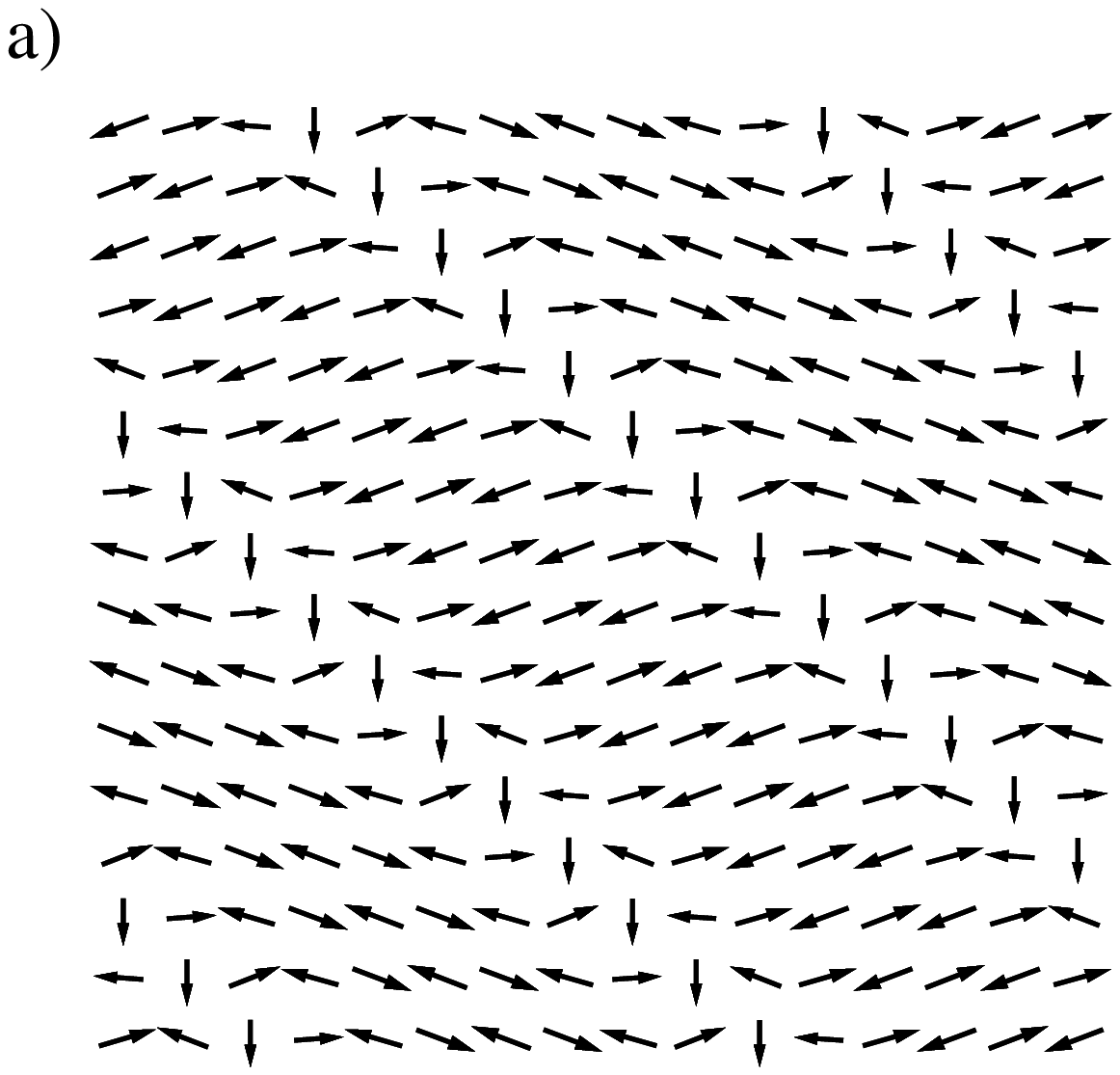}
\includegraphics[width=9cm,clip=true]{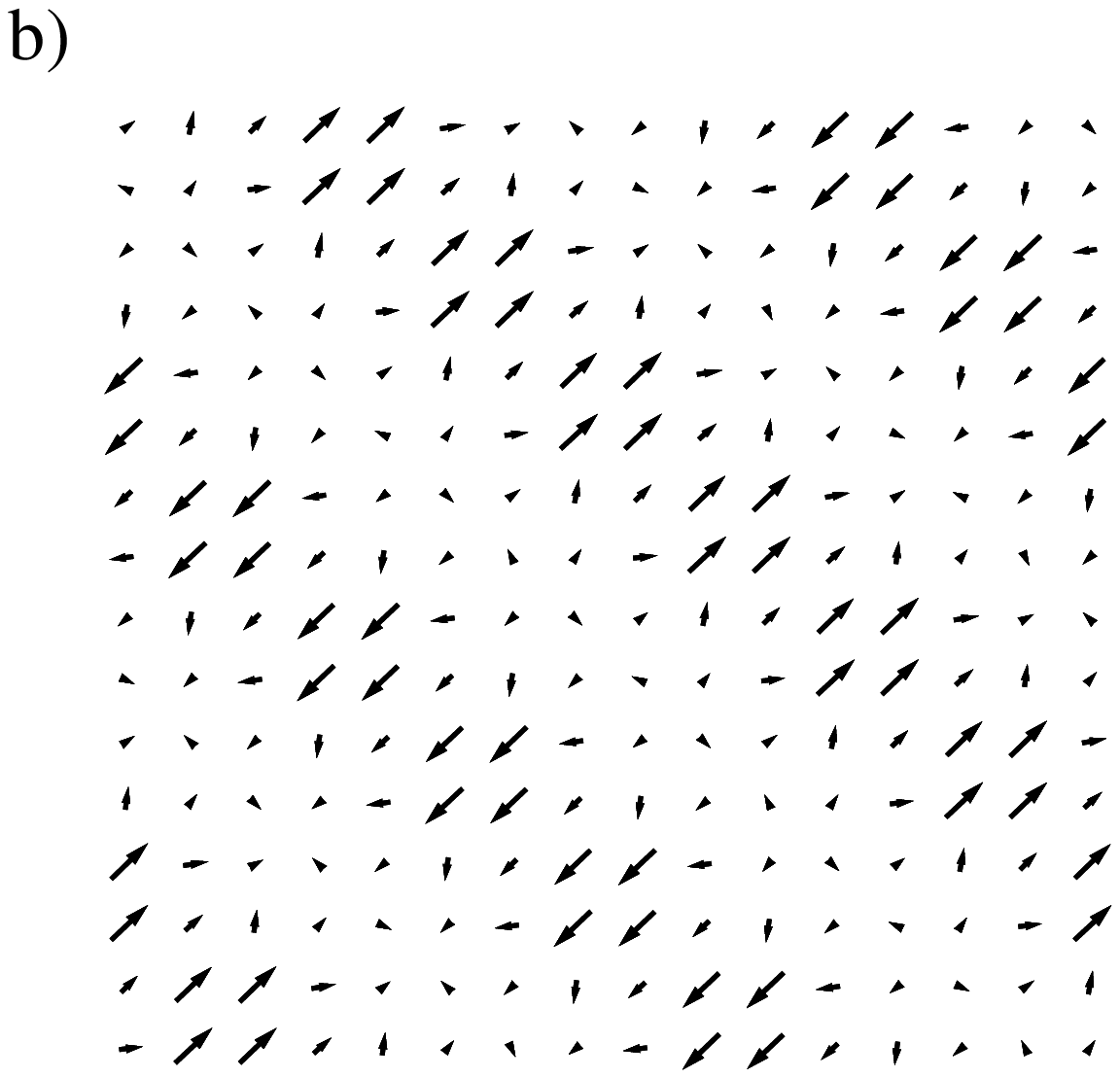}
\caption{a) Diagonal stripe solution with spin canting. The 
(hole) charge is accumulated on the lines with ferromagnetically
aligned spins (parallel to the z-direction). b) Corresponding
pattern of DM spin-currents ${\bf j}^{DM}_i$ which are assigned to the lattice
by summing the in- and outgoing currents of the connecting bonds, i.e.
${\bf j}^{DM}_i=\sum_{j}{\bf j}^{DM}_{ij}$.
$16\times 16$ system with 16 holes corresponding
to doping $x=1/16$. $U/t=8$, $t'/t=-0.4$.} 
\label{fig3}
\end{figure}
The above finding that low doping spirals are unstable towards
phase separation does not require that macroscopic phase separation is
the ground state. There could exist inhomogeneous solutions with
even lower energy. One could envisage e.g. an elliptical spiral
\cite{zachar} corresponding to spin structure
\begin{eqnarray}
S_i^x &=& S_0 \cos(\alpha)\cos(\Qvec\Rvec_i) \nonumber \\
S_i^y &=& S_0 \sin(\alpha)\sin(\Qvec\Rvec_i).  \label{eq:ellipt}
\end{eqnarray}
which connects stripe and spiral states [cf. Eq. (\ref{eq:spiral})] 
by tuning the eccentricity from $\alpha=0$ to $\alpha=\pi/4$.
As discussed below there are more complex structures which minimize
the energy. 

By performing an unconstrained minimization of the spin-rotational
GA energy functional \cite{goe98}  Eq. (\ref{ega}) on finite clusters, 
we find that for $t'/t=-0.4$
the low doping solutions are diagonal stripes with significant
spin canting. An example is shown in Fig. \ref{fig3}(a) and the corresponding
energy is indicated by the full square in Fig. \ref{fig2}(b).
This texture is characterized by a domain wall of the
antiferromagnetic order which has a fractional phase change of the AF
order parameter $\Delta \theta < \pi$ (with $\theta$ the angle
between the staggered magnetization and the quantization axis) 
contrary to collinear stripes which have $\Delta \theta =\pi$.
Thus instead of building up a macroscopic phase separation
between undoped AF and doped spiral regions (doping
$\delta \approx 0.11$ as can be read from Fig. \ref{fig2}b), 
the system prefers to separate these textures 
at the nanoscale, corresponding to the
modulated structure shown in Fig. \ref{fig3}(a). 
The material has clearly separated into ferromagnetically ordered diagonal 
lines which are at the center 
of 3-site wide spiral stripes with an average doping of $\delta \approx 0.11$
and  undoped AF regions located between these stripes.

Research on multiferroics\cite{Cheong,Katsnelson} has shown that spin
canting produces a force on the ligand ions via the  Dzyaloshinskii-Moriya (DM)
interaction~\cite{Dzyaloshinskii,Moriya,Bonesteel,Koshibae}
 which in some cases can lead to a uniform dipole
moment. For the complicate textures we find the distortion is not
uniform but can act as an experimental signature of the canted phases
which we describe in some detail.  

If we think of the one-band Hubbard model 
as embedded in the cuprate lattice, then the spin canting causes a force on the oxygen atoms, leading to a structural distortion of the CuO$_2$ planes.  
In the presence of spin-orbit interactions one can add to the Hamiltonian 
the following
DM and elastic term:  
\begin{eqnarray}
  \label{eq:h1d}
H_{DM}&=&\sum_n \lambda {\bf u}_{n+1/2} \cdot {\bf f}^{DM}_{n,n+1},\\
H_{E}&=&\sum_n\frac{k}2|{\bf u}_{n+1/2} |^2
\nonumber
\end{eqnarray}
where ${\bf u}$ are the oxygen displacements and the DM force field
is given by 
${\bf f}^{DM}_{ij}\equiv {\bf e}_{i,j}\times {\bf S}_i\times {\bf S}_j$. 
Here ${\bf e}_{n,n+1}$ is a unit vector joining nearest neighbor Cu
atoms.  The product ${\bf S}_i\times {\bf S}_j$
is also known as the chiral vector order parameter.\cite{one07} 
 
For definiteness, we assume that the spins lie in the CuO$_2$ plane, as is 
approximately true for the cuprates. In this case, only the components of 
${\bf S}_i\times {\bf S}_j$
perpendicular to this plane are finite, and ${\bf f}^{DM}_{ij}$ lies in the 
plane at right angles to the Cu-O-Cu bond.  Once ${\bf f}^{DM}_{ij}$ becomes 
finite, $H_E+H_{DM}$ is minimized by developing a finite in-plane oxygen 
displacement perpendicular to
the Cu-Cu bond. Due to the breaking of inversion symmetry around the
oxygen atoms, this relaxation creates a local dipole moment.

In the following we use the `DM current' (or force flux)
\begin{equation}
{\bf j}^{DM}_{ij}\equiv |{\bf f}^{DM}_{ij}| {\bf e}_{i,j}
=|{\bf S}_i\times {\bf S}_j|{\bf e}_{i,j}
\end{equation}
in order to characterize the ground state magnetization 
pattern.~\cite{note,schuetz04} These currents lie in the same plane as the spins and
for the canted diagonal stripes are shown in
Fig. \ref{fig3}(b). One observes a maximal current perpendicular to the
stripes with its direction alternating from stripe
to stripe, so that the periodicity is the same as for the underlying
spin structure.  In contrast to the elliptic spirals, Eq. (\ref{eq:ellipt}), 
which have a finite net force flux, the total spin force vanishes
for our diagonal solution with an even number of stripes. 
The structure of the texture shown 
in Fig. \ref{fig3} can be decomposed in harmonics as
\begin{eqnarray}
S_i^x &=&\sum_n S_n^x \cos(\Qvec^{(n)}\Rvec_i) \nonumber \\
S_i^y &=&\sum_n S_n^y \cos(\Qvec^{(n)}\Rvec_i) 
\label{eq:ellipt2}
\end{eqnarray}
with ${\bf q}=(\frac{2\pi}{16},\frac{2\pi}{16})$ and we have 
set $\Qvec^{(n)}=\Qvec_{AF}-n\qvec$. The solution breaks spin
rotational symmetry, so the relative weights of the Fourier
components on the $x$ and $y$ magnetization directions depend on the
particular solution or quantization axis. For the choice
shown in Fig.~\ref{fig3} the amplitudes are zero in the $x$ ($y$)
direction for $n$ even (odd).  The nonzero amplitudes are given
by $S_1^x\approx 0.24$, $S_0^y\approx0.03$, and $S_2^y\approx0.1$. 
Therefore the x-component of the spin structure [Fig. \ref{fig3}(a)] 
induces incommensurate correlations at 
${\Qvec}=(\pi-\frac{2\pi}{16},\pi-\frac{2\pi}{16})$ and weaker ones at 
higher harmonics,
${\Qvec}=(\pi-\frac{4\pi}{16},\pi-\frac{4\pi}{16})$, etc. 
In addition, the y-component leads to (weaker)  
commensurate correlations at ${\Qvec}=(\pi,\pi)$ and 
${\Qvec}=(\pi-\frac{3\pi}{16},
\pi-\frac{3\pi}{16})$, etc.. Interestingly, this implies that in neutron
scattering a weak commensurate peak will appear, a feature that may be
hard to distinguish experimentally from phase separation between
commensurate and incommensurate stripes.  

For smaller values of $|t'/t|$, stripes do not profit from
the transverse spin degrees of freedom and the energy of the spin canted
solution is approximately that for the collinear solutions (cf. square
symbol in Fig. \ref{fig2}(a)). 
Nevertheless, the finding that very different patterns are so close 
in energy suggests that stripes will be very susceptible to quenched
disorder, inducing charge-spin glass behavior and making it difficult to
obtain clear stripes signatures. 

\subsection{Checkerboard order for larger $|t'/t|$}
\begin{figure}[thb]
\includegraphics[width=9cm,clip=true]{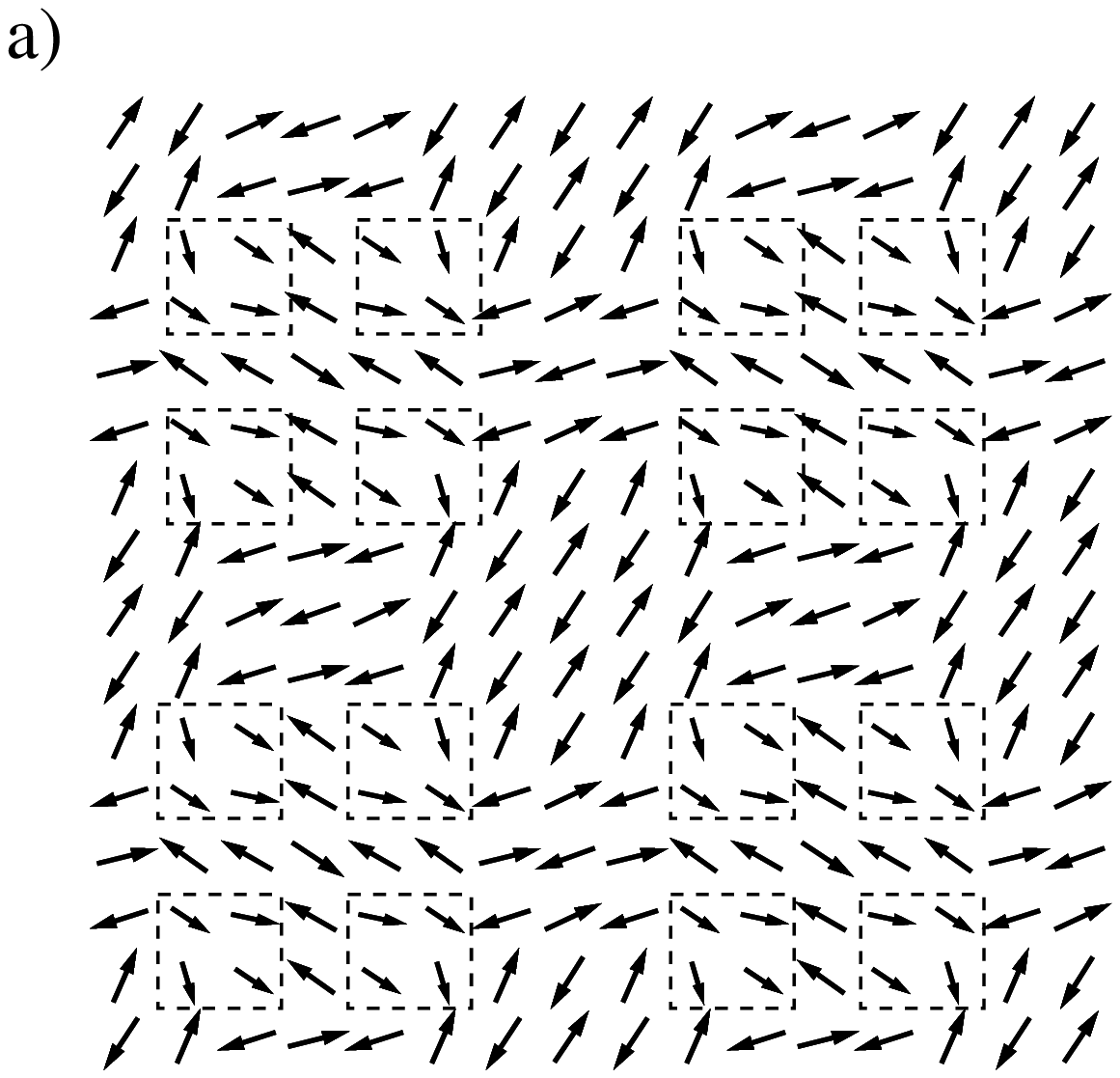}
\includegraphics[width=9cm,clip=true]{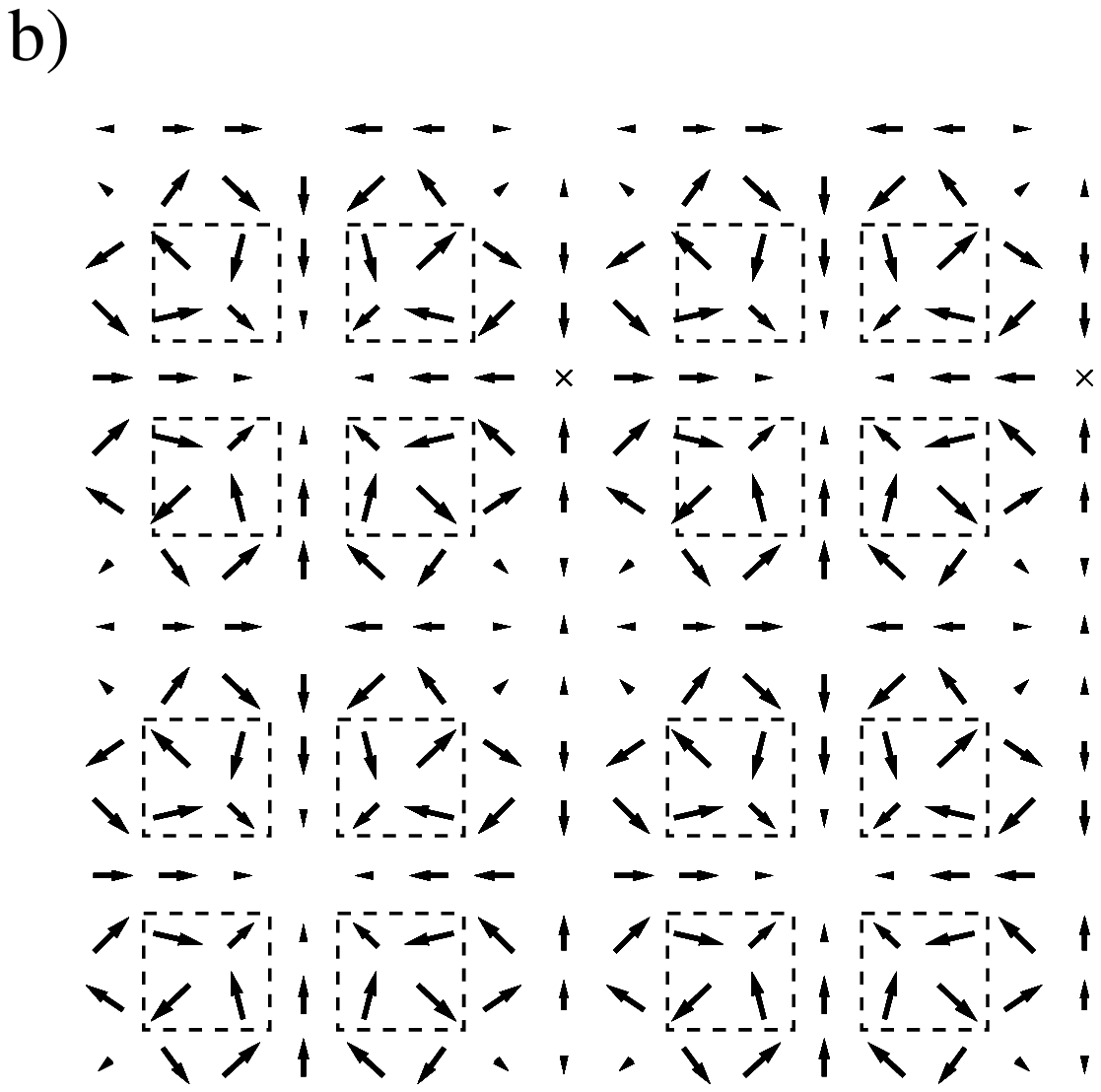}
\caption{Spin structure (a) and DM spin currents (b) for the two 
dimensional `checkerboard' structure with spin canting. Regions
with larger hole density are enclosed by boxes. 
$16\times 16$ system with 16 holes corresponding
to doping $x=1/16$. $U/t=8$, $t'/t=-0.5$.} 
\label{fig4}
\end{figure}
In general, the spin canting allows the system to make a compromise between
ferromagnetic order (which is favored for large $|t'/t|$ \cite{toh94,han97}
and AF order induced by the local correlations. 
The fact that local ferromagnetism occurs for large values of the next-nearest
neighbor hopping was also found to play a role in the  
stabilization of two-dimensional checkerboard textures
with small ferron type clusters for large values of $t'/t=-0.5$.~\cite{sei07}
These structures optimize the gain in kinetic
energy for a collinear two-dimensional spin arrangement and turn out to
be more stable than (collinear) stripes in this parameter regime.
Based on our present findings it is thus natural to investigate the
stability of spin-canted checkerboard textures. 

Fig. \ref{fig4}(a) shows the minimum energy solution for $16$ holes
on a $16 \times 16$ lattice for $t'/t=-0.5$. Similar to our
previous findings in Ref. \onlinecite{sei07} it is only
for such large values of $t'/t$ that two-dimensional textures
are stabilized over the (spin-canted) stripe solutions.
The doped holes are confined to $2\times 2$ plaquettes which are
indicated by the dashed squares in Fig. \ref{fig4}(a).
The overall pattern consists of a `checkerboard' array of boxes each 
containing four charge plaquettes.
The planar spin structure is no longer incommensurate but has
dominant correlations at the AF wave-vector together with a finite
ferromagnetic moment. Again, the spin structure can be characterized
by the associated ``DM spin currents'' which are shown in
Fig. \ref{fig4}(b). Here the centers of each box appear as    
 a spin current sink whereas the center and corners of the lattice between boxes
appear as sources for the spin current. 
Note that: 1) there is no net DM spin current as for a spiral so there will be
 no net dipole moment, 2) according to our definition  the DM spin
 currents are useful to visualize the expected lattice distortions but
 do not correspond to currents of a conserved quantity. 

We have also investigated the stability of  vortex lattices
which have been proposed by Fine \cite{fine07} but found them
always higher in energy than the checkerboard-type structures. 

\section{Discussion and Conclusions}

In a recent publication,~\cite{bob1} the magnetic
phase diagram of cuprates was computed in terms of Fermi surface
nesting, which determined the  
critical interactions $U$ as a function of doping, and the leading incommensurate $\Qvec$ 
vectors at threshold.  The present calculations work at fixed $U$ above threshold, but in 
general find very similar $\Qvec$ values at fixed doping.  Thus, for $t'/t=-0.4$, Fig.~3(b) 
shows a crossover from diagonal to vertical $\Qvec$ as doping $\delta$ decreases.  While 
Ref.~\onlinecite{bob1} correctly predicted this {\it local} minimum, it missed the fact 
that the {\it global} minimum is determined by (nanoscale) phase separation.  Similarly 
for $t'/t=-0.2$, the $\Qvec$ of the vertical stripes is consistent with Fermi surface 
nesting\cite{bob1} for $\delta \ge 1/8$, but not at lower doping.

Our investigations have revealed that the nature of symmetry-broken states
in underdoped cuprates strongly depends on the strength of the next-nearest
neighbor hopping. A value of $t'/t$, which is appropriate for lanthanum
cuprates \cite{pava} stabilizes striped ground states so that the
present analysis supports our related calculations on charge and magnetic 
excitations \cite{lor03,sei05,sei06} as well as transport 
properties \cite{lor02} in these compounds.
Naturally the wave-function of such inhomogeneous textures is determined 
by states far from the Fermi energy so that one may question the present single-band analysis. 
However, we have checked by explicitly evaluating spirals and stripes
within the three-band model that the conclusions of the present
paper stay valid but the role of $t'/t$ is played by the oxygen-oxygen hopping
parameter.

In Fig. \ref{fig1b} we also show data from recent neutron scattering
experiments \cite{he10} on Fe-LSCO. In these overdoped samples an elastic 
incommensurate spin response was found close to the dominant nesting vectors
as extracted from ARPES experiments. It was thus concluded that the induced 
incommensurate response signals an inherent instability of the itinerant
charge carriers, being different from the low doping stripes arising
from localized Cu spins.
Within our calculations (cf. Fig. \ref{fig1b}) the data are rather close
to the incommensurability curve of stripes obtained for $t'/t=-0.2$, which
is the appropriate value for LSCO.

 Since our calculations are based on an
itinerant approach, there is no `dichotomy' between low doping `localized'
spin stripes and large doping itinerant ones, but both appear as different
limits of the same model. In fact, at these large dopings
one can also regard the stripes as arising from an instability of the
`nearby' (in energy) paramagnet similar to the spiral case as discussed
in context to Fig. \ref{fig1}. In other words, increasing the doping
one can go continuously from a stripe state at low doping to an
itinerant spin-density wave (SDW) at high doping which disappears at a
second order transition to become a uniform paramagnet.

 At fixed doping one finds an analogous SDW instability with increasing $U$. 
For $t'/t=-0.2$ and $\delta=0.3$  
the spin density wave instability occurs at $\Qvec = (0.67\pi,\pi)$ and 
critical interaction $U_c/t\sim 2$. We can therefore view the
$d=3$ stripes (which have a similar periodicity) as the realization
of this instability whereas spirals for these parameters have higher 
energy. Our  GA calculations 
yield  a charge modulation of only $8\%$ for the $d=3$ stripes at
$\delta \sim 0.25$ 
which  should be  considered as an upper bound since our calculation 
 neglects the effect of fluctuations. 
It may be therefore experimentally difficult
to distinguish between a stripe and spiral state in this doping range.

For larger values of $|t'/t|$  as relevant for e.g. bismuthate 
high-T$_c$ cuprates we have 
found that spiral textures are lower in energy than 
stripes in which canting is not allowed. If this latter restriction is
relaxed, stripes and checkerboard states with a substantial
spin canting become again more stable. 
This can be understood from the fact that spirals are 
unstable at low doping towards phase separation so that stripes and
checkerboard states can be seen as nanoscale phase separated states.

It is interesting to remark that the more stable stripe at small
doping is a novel kind of stripe with a fractional change in the phase
of the order parameter as opposed to the usual $\pi$ change. These
states can be experimentally detected because they have a finite Bragg peak
weight both at the commensurate AF wave vector $\Qvec_{AF}={\pi,\pi}$
and at the incommensurate positions but can also be easily  confused
with phase  separation among stripes and the commensurate
antiferromagnet phase.

 As shown in Figs. \ref{fig3}(b) and \ref{fig4}(b) the spin canting
induces the flow of DM spin currents along the bonds of the lattice
which are associated with the force flux ${\bf f}^{DM}_{ij}$.
Due to the breaking 
of inversion symmetry arising from the oxygen displacements,
such DM spin currents may be associated with an electric polarization 
${\bf P}\sim {\bf f}^{DM}_{ij}$.~\cite{katsura05}

For the diagonal structure shown in Fig. \ref{fig3}(b), ${\bf P}$ would point
along the stripe but in alternate directions from stripe to stripe.
On the other hand, in the nearly AF ordered regions between 
the domain walls the DM spin currents and thus the electric polarization 
almost vanish.
The spin canted stripe can therefore also be considered as a one-dimensional 
`antiferroelectric' with a vanishing net polarization.~\cite{note2}  

In the overdoped regime and for sufficiently large $|t'/t|$ spirals do
not phase separate and under certain conditions may constitute 
the ground state in cuprate superconductors. In this regard, the recent
elastic NS study \cite{hiraka10} on  
(Bi,Pb)2201 codoped with iron is interesting
since it reveals incommensurate magnetic correlations with
$\varepsilon\approx 0.21$ at a doping concentration $\delta \approx 0.23$.
Our present computations shown in Fig. \ref{fig1b} suggest that the 
elastic neutron scattering data from Ref. \onlinecite{hiraka10} 
are compatible with vertical
spirals provided that the value of the next-nearest hopping
for Bi2201 is in the range  $-0.4 < t'/t < -0.2$. LDA calculations by
Pavarini and coworkers \cite{pava} yield a value of $t'/t\approx -0.25$
for Bi$_2$Sr$_2$CuO$_6$ where our results (cf. Fig. \ref{fig2}) suggest that
vertical spirals are still more favorable than diagonal ones and may even
dominate over stripes.
Naturally, our computations which are for perfectly modulated spirals, can
only be qualitative since the measured coherence length of the magnetic 
modulation \cite{hiraka10} is of the order of the iron distance.
Therefore additional effects like disorder and the magnetic moments
arising from the Fe dopants should also be taken into account which
is, however, beyond the scope of the present paper. 
 
In the underdoped regime it is unlikely that uniform spirals will
survive. Although for large $|t'/t|$ the phase separated spiral has
nominally lower energy than the stripes, the macroscopically  
phase separated state
implies a charge imbalance $\sim 0.1e $ per Cu which would result in
a prohibitive Coulomb cost.\cite{ort06} Even without long-range Coulomb
interaction a nanoscale phase separated state has lower
energy. 

Clearly the spin polarons, the checkerboard states as well as
stripes, having charge inhomogeneities, will couple with impurities
resulting in a spin and  charge spin glass state consistent with
experiments. Since this state will have substantial spin-canting it
becomes rather semantic to distinguish it from the disordered spiral
ground state as proposed by Sushkov.~\cite{sush05,sush07,sush09} For
small $|t'/t|$ we expect that short-range diagonal stripes order
dominates in this state as observed experimentally although with a
correlation length of the static incommensurate response 
of the same order as the periodicity.~\cite{waki00} 
One should therefore rather think of a disordered state 
with meandering stripe  which can profit from the impurity potential.

\acknowledgements
G.S. acknowledges financial support from the Deutsche Forschungsgemeinschaft.
RSM's  research is supported by the U.S.D.O.E contracts
DE-FG02-07ER46352 and DE-AC03-76SF00098. J. Lorenzana's research is
partially supported by the Italian Institute of Technology-Seed
project NEWDFESCM.


\begin{thebibliography}{10}
\bibitem{note2} The `antiferroelectric' situation depends of course 
                on the periodicity of the stripe structure. A different 
                periodicity along $(1,1)$ may also result in a ferroelectric
                state.
\end{thebibliography}
\end{document}